\documentclass[aps,prl,twocolumn,superscriptaddress,showpacs]{revtex4}
\usepackage{graphicx}
\usepackage{bm}
\usepackage{color}
\input colordvi
\begin{document}
\title{Integrable Mott insulators driven by finite electric field}
\author {Marcin Mierzejewski}
\affiliation{Institute of Physics, University of Silesia, 40-007 Katowice, Poland}
\author{Janez Bon\v ca}
\affiliation{Faculty of Mathematics and Physics, University of Ljubljana, SI-1000 Ljubljana, Slovenia }
\affiliation{J. Stefan Institute, SI-1000 Ljubljana, Slovenia }
\author{Peter Prelov\v{s}ek}
\affiliation{Faculty of Mathematics and Physics, University of Ljubljana, SI-1000 Ljubljana, Slovenia }
\affiliation{J. Stefan Institute, SI-1000 Ljubljana, Slovenia }

\begin{abstract}
We develop a  method for extracting the steady nonequilibrium current
from studies of driven isolated systems, applying it to the model of
one-dimensional Mott insulator at high temperatures. While in the
nonintegrable model the nonequilibrium conditions can be accounted by internal heating,
the integrability leads to a strongly nonlinear dc response
with vanishingly small dc conductivity in the linear--response
regime. The finding is consistent with equilibrium results for dc limit of
the optical conductivity determined in the presence of a weak and
decreasing perturbation.
\end{abstract}
% 71.27.+a Strongly correlated electron systems; heavy fermions
% 72.10.-d Theory of electronic transport; scattering mechanisms
% 72.10.Bg General formulation of transport theory  

\maketitle

\textit{Introduction.}--- 
Understanding the differences between integrable and nonintegrable systems 
has been recognized  as a mostly theoretical problem.
However, recent advances in the physics of ultracold atoms enable 
direct experimental verification of the underlying theoretical concepts.  
%Certainly, 
The presence of conserved quantities (a hallmark of integrability) 
restricts thermalization processes \cite{kollar}, hence integrable systems may
relax to the generalized Gibbs state \cite{rigol} instead of the
thermal state. Qualitative differences between both system types show up also 
in their linear--response (LR) to an external field $F$. 
Although integrability is usually broken by $F$, this breaking is not visible in the LR theory,
since  LR susceptibility is calculated at $F=0$. The role of integrability 
at finite driving remains important but largely unexplored field. 
Here we study insulators driven by electric field $F$ and investigate
how the integrability--related properties of the LR regime decay/change with increasing driving.

In the integrable {\em metallic} systems there is a dissipationless 
component of the LR optical conductivity  $\sigma(\omega)$ at arbitrary temperature 
$T$ \cite{u2}. 
%Since this component is proportional to $\delta(\omega)$, it is visible in the LR response at arbitrary time--scale.
The difference between integrable and nonintegrable metals persists also beyond the LR regime but only
for restricted time of driving \cite{prl1}. 
The response of integrable systems is dominated by the Bloch oscillations which, however, 
are damped due to field--induced breaking of integrability. 

In case of Mott insulators at  $T>0$,  
LR 
%to driving by electric field 
has proven to represent 
by itself a theoretical challenge not fully understood so far, 
with a transport fundamentally different from a 
usual picture of a band insulator at $T>0$. 
It is plausible that generic nonintegrable models do not show dissipationless
transport at $T>0$ as characterized by the charge stiffness
$D(T)>0$.  This seems to be quite settled also for some prototype one--dimensional 
integrable models within the insulating regime \cite{zotos,afleck}, as e.g. the
$t$--$V$  (or anisotropic Heisenberg) model.
However, in the latter case also the regular dynamical conductivity 
$\sigma(\omega \to 0)$ vanishes in finite systems  at $T>0$ \cite{u4}
(stimulating the scenario of an 'ideal insulator' \cite{u2}), but $\sigma(\omega)$
shows at the same time anomalously large finite--size
contribution at $\omega \propto 1/L$  which at the first sight might 
be consistent with thermodynamic value $\sigma(0) >0$.   
It should be noted that $\sigma(0) >0$  may appear as a result of steady-current
calculations for an insulator with attached leads \cite{znid} 
as well as of the  time--evolution analysis \cite{stein}. 
In view of these open questions \cite{sah}, it is quite unclear which might be the effect of finite but  weak driving 
$ F>0$ as well as of  small nonintegrable perturbation. 

In this Letter we show that for nonintegrable insulator one can indeed reconcile the response 
to finite--field driving 
with the LR result if we take into account the energy increase (the Joule heating) and the existence of 
(time-dependent) quasi-equilibrium after a short-time transient. More dramatic are  results for driven integrable insulators.  Since  $F \ne 0$ breaks integrability large $F$--response is quite similar to 
the nonintegrable case. On the other hand, our results for weak field $F \to 0$ indicate the vanishing of the 
steady current response $\sigma (0)= I/F$ reviving the claim of an 'ideal insulator' with $\sigma (0) \to 0$
at $T>0$. Such scenario seems to follow also from the alternative approach 
calculating LR $\sigma(0)$ with a weak and decreasing 
integrability--breaking perturbation.

\textit{Model.}--- 
A one--dimensional system of interacting spinless 
fermions  with periodic boundary  conditions is threaded
by time--dependent magnetic flux $\phi(t)$
\begin{eqnarray} 
H & = &-t_h \sum_j \left\{ {\mathrm e}^{i \phi(t)}\; c^{\dagger}_{j+1}c_j +{\mathrm h.c.} \right\} \nonumber \\
&& + V \sum_j \hat{n_j} \hat{n}_{j+1} +W \sum_j \hat{n_j} \hat{n}_{j+2} \label{ham},
\end{eqnarray}
where  $\hat{n_j}=c^{\dagger}_{j}c_j $, $t_h$ is the nearest--neighbor (nn) hopping integral,
$V$ and $W$ are the nn and next nn repulsive potentials, respectively.    In the absence of external driving 
the system is integrable for $W=0$, whereas  nonvanishing $W$ breaks the integrability. 
As we investigate the physics of Mott insulators, we consider a half--filled system with $L$ sites and $L/2$ fermions.
Initially $\phi(t=0)=0$ and the system is assumed to be in an (approximate) {\em microcanonical} state. 
Then, for $t>0$  the flux $\phi(t)=-Ft$ gives rise to a constant electric field $F$. 
The field will be expressed in units of $[t_h/e_0a]$, where $e_0$ is the unit charge and $a$ is the lattice distance.  We set $t_h=\hbar=e_0=a=1$.
The charge current  $I(t)$  can be calculated either  directly from the expectation value of the current operator:
\begin{equation}
\hat{J}=\frac{i}{L} \sum_j \left\{  {\mathrm e}^{i  \phi(t)}  \; c^{\dagger}_{j+1}c_j -{\mathrm h.c.} \right\},
\label{jdef}
\end{equation} 
or from the increase of the total energy \cite{prl1}
\begin{equation}
 \dot{E}(t) = \frac{\mathrm{d}}{\mathrm{d} t} \langle H(t) \rangle= L\;F(t)\;I(t). \label{hdot}  
\end{equation}

\begin{figure}
\includegraphics[width=0.48\textwidth]{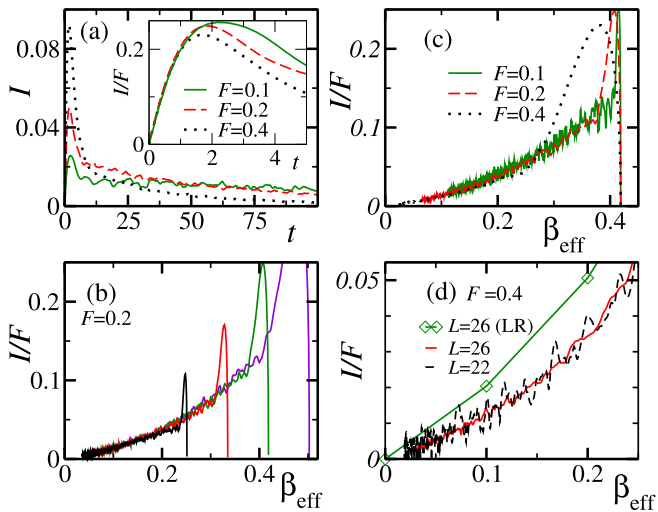}
\caption{(Color online) Nonintegrable case $V=3, W=1$, $L=26$ with 
a pure initial state ($N_s=1$).
(a): current $I(t)$ (main) and $I(t)/F$ (inset) for initial $\beta_{eff} \simeq 0.42$;
(b) and (c): $I/F$ vs. $\beta_{eff}$ for various initial energies $E_0$ (panel b) 
and various $F$ (panel c);
(d): The same as in c) but for various $L$ and compared with the results of equilibrium
LR theory.}
\label{fig1}
\end{figure}

Calculating  the dc current $I$ for finite electric field is a challenging problem.  
Literally, the dc response (steady nonequilibrium state) is expected  only for  open quantum systems 
when the energy gained by carries moving under $F$ [see Eq. (\ref{hdot})] 
is transferred to  the heat reservoirs.  
In driven isolated systems however, the energy steadily increases causing a steady decay 
of $I(t)$. 
This holds true independently of integrability or the system size and remains valid also in the
thermodynamic limit.
We first analyze nonintegrable systems and show how $I$ can be reliably 
determined from the decaying  $I(t)$. 

\textit{Nonintegrable case.} ---As the technical aspects of our computational approach are 
described in Ref. \cite{prl1}, here we recall only the basic steps.   
We apply microcanonical Lanczos method \cite{mclm} and numerically generate $N_s$ initial states $|\Psi_l(0) \rangle$
with assumed energy $E_0= \langle H(t=0) \rangle$  but as well with a small energy uncertainty $\delta E=\langle [H(0)-E_0]^2 \rangle^{1/2} \sim 10^{-2}$. 
Then, the time evolution of $|\Psi_l(t) \rangle$ is calculated by step--vise change of the flux $\phi(t)$ in small 
time increments $\delta t \ll 1$ applying  the Lanczos time propagation method \cite{lantime} to each interval $\delta t$.
 The larger the system is, the fewer initial states are needed. It will be shown that for large $L$
%one can even start from a pure initial state ($N_s=1$) and
 one can take $N_s=1$ and omit the averaging over $|\Psi_l(0) \rangle$. 
Similar observation  has been reported in Ref. \cite{rigolnature}
but for relaxation 
instead of evolution upon driving.

Inset in Fig.  \ref{fig1}a shows the ratio $I(t)/F$ for a nonintegrable system.  
After a relatively short time of driving  ($t\sim 2 $) results for different fields split marking the end of 
the equilibrium LR regime. This nonlinearity becomes even more important for longer driving, when the 
largest currents occur for the  weakest fields (see main panel). In the following we 
show that such behavior occurs simply due to heating. 
For this sake it is convenient to introduce a renormalized 
energy $\varepsilon(t)=[E_{\infty}-E(t)]/L$, where 
\begin{equation} 
E_{\infty}=L(V+W)\frac{L/2-1}{2(L-1)},
\end{equation}
denotes the energy for $T \to \infty$. In equilibrium and at high temperatures
$\varepsilon$ is proportional to the inverse temperature $\beta$ 
\begin{equation}
\frac{\varepsilon}{\beta}=\frac{1}{2}+\frac{V^2+W^2}{16}. 
\label{beff}
\end{equation}
We use an effective inverse temperature
$\beta_{eff}$ defined by Eq. (\ref{beff}) also for driven systems, when
this quantity should be understood {\em only} as a 
measure of the instantaneous energy. Upon driving $\beta_{eff}(t)$ monotonically decreases provided $I(t)$ does not change sign. 
%Then, there is one--to--one correspondence  between $\beta_{eff}$ and $t$. 

Fig. \ref{fig1}c shows the same data as panel (a) but the ratio
$I/F$ is now plotted as a function of $\beta_{eff}(t)$ instead of $t$.  
After the system passes the transient state with a characteristic peak in $I(t)$, 
results for different $F$ nearly overlap.
Hence, the proportionality between $I$ and $F$ is restored 
far beyond the LR regime provided that 
the effect of the Joule heating is appropriately subtracted.
It also explains the origin of nonlinearity visible in Fig. \ref{fig1}a: 
fixing the time of driving while allowing for different $F$ one actually 
compares systems with exceedingly different energies. 
Such inconsistency is not a finite--size effect 
but is generic for any isolated system with a non--vanishing carrier density.
As shown in Fig. $\ref{fig1}b$, $I(\beta_{eff})/F$ is almost independent 
of the initial energy $E_0$ or the choice of $| \Psi_l (0) \rangle $. 
This ratio is uniquely determined by the instantaneous energy ($\beta_{eff}$)
up to fluctuations which vanish in the thermodynamic limit (see panel d).  
Fig. \ref{fig1}d  also demonstrates that $I[\beta_{eff}(t)]$ 
is close to the the  dc current obtained from the equilibrium LR as described in
more detail later.

Result presented in Fig. \ref{fig1} consistently support the following picture:
the isolated system evolves upon finite $F$ towards the infinite--temperature state 
but the (quasi) dc current is uniquely determined by the instantaneous energy,
suggesting a quasi--equilibrium evolution.
Attaching the system to a heat reservoir should set $\beta_{eff}$ 
leading to $I[\beta_{eff}(t)]=const$.
However, in such case $I$ could depend on the details of
the coupling to reservoir.
This observation allows
us to estimate intrinsic  $I$  for finite $F$ even though  calculations
are carried out on  an isolated system. As expected for nonintegrable insulators, 
the  Joule heating represents the central  source of nonlinearity at high temperatures.

 \begin{figure}
\includegraphics[width=0.48\textwidth]{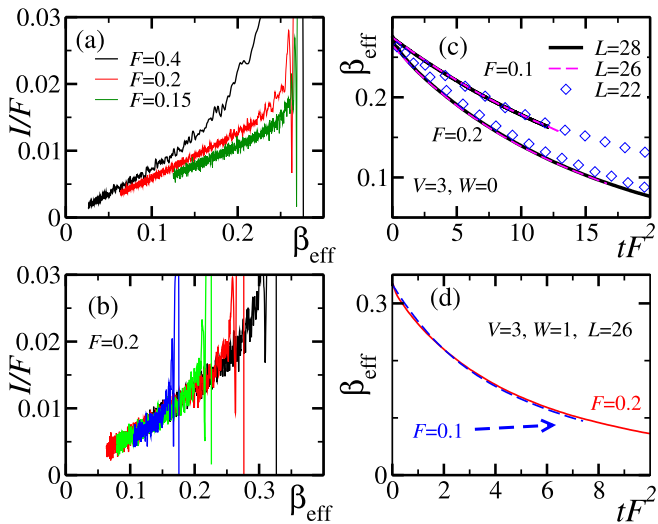}
\caption{(Color online) 
(a) and (b):  $I/F$ vs. $\beta_{eff}$ for 
integrable system $V=3, W=0$ and $L=28$.
(c) and (d) $\beta_{eff}$  vs. $t F^2$ for
integrable (c)
and nonintegrable (d) systems.
In (c) curves are shifted horizontally so they
overlap for large times. $N_s=1$ apart from  $F=0.15$ and $F=0.2$ 
in (a) where $N_s=7$. 
}
\label{fig2}
\end{figure}

\begin{figure}
\includegraphics[width=0.48\textwidth]{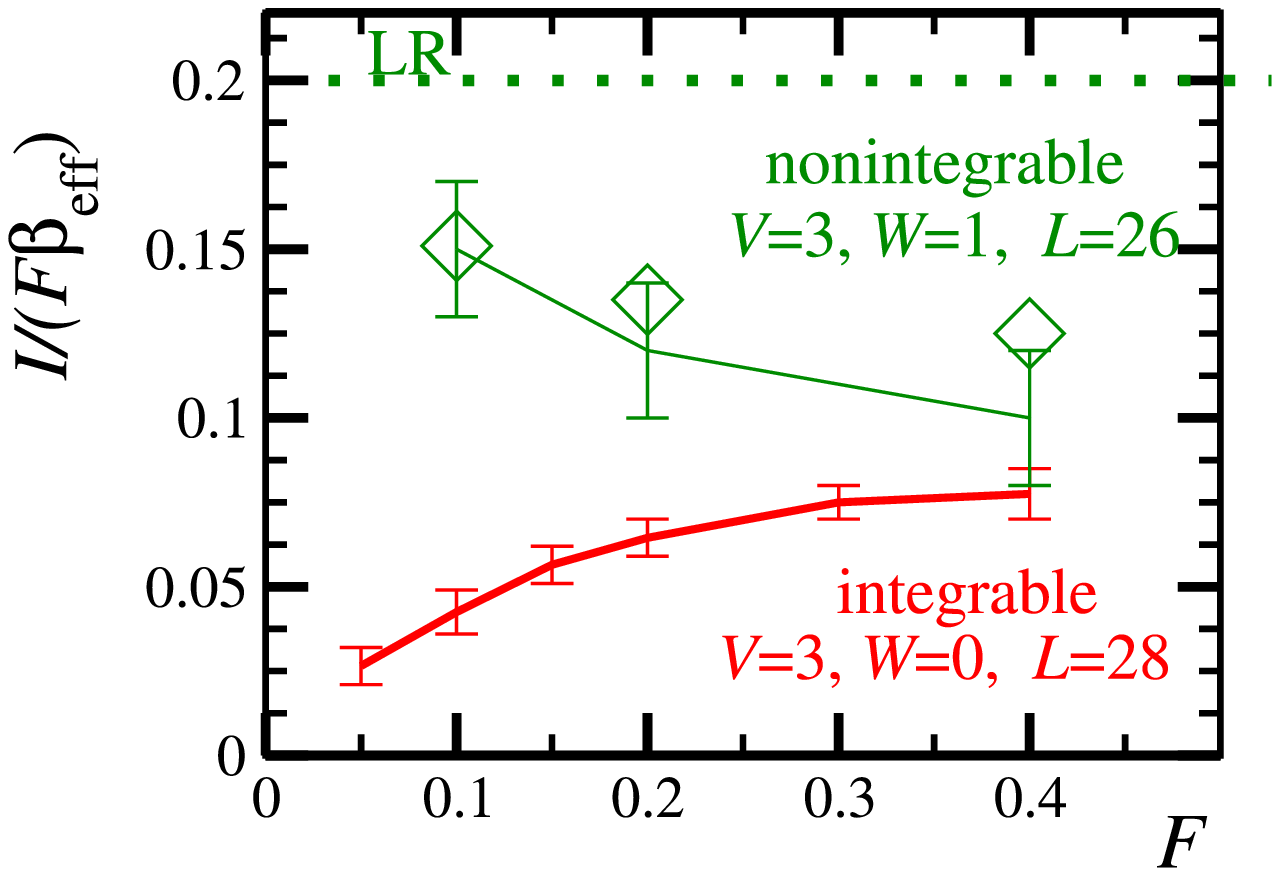}
\caption{(Color online) $I/(\beta_{eff} F)$ estimated from 
$E(t)$ [Eq. (\ref{hdot})] for $\beta_{eff} \rightarrow 0 $ (points with guidelines).
For nonintegrable case results for
$\beta_{eff}=0.1$ (diamonds)  together with LR  for $\beta=0.1$ 
(dotted line) are shown as well. }
\label{fig2_extra}
\end{figure}

\textit{Integrable case.} ---Before applying the same approach to the integrable insulators
one should address the question whether $I$ can be again uniquely determined (up to finite--size
fluctuations) by the instantaneous energy. Since finite $F$ breaks integrability such scenario
is possible and as follows from  Fig. \ref{fig2}b it actually takes place. 
Although,  for a pure initial state ($N_s=1$) the finite--size fluctuations are much larger 
than in the nonintegrable case (Fig. \ref{fig1}c), $I(\beta_{eff})$ for 
various initial states do  overlap.  Therefore, one can apply the same method of reasoning as for nonintegrable insulators
and  take into account the heating (the energy increase). 
The results are shown in Fig. \ref{fig2}a. Here, one
finds important qualitative differences between both system types.
While in the nonintegrable 
case the ratio $I/F$ slightly decreases with $F$ (see Fig. \ref{fig1}c),
now an opposite dependence is visible.
Namely, $I(\beta_{eff})/F$ (strongly) increases with $F$ indicating on
a nonlinear mechanism  beyond that of the Joule heating.  
%This nonlinear effect is  even more pronounced (since fast fluctuations of  $I(t)$  are averaged) in $E(t)$ or equivalently $\beta_{eff}(t)$. 
If the dependence between 
$F$ and $I$ were linear [$I(E,F)/F=\gamma(E)$], one would obtain
from Eq. (\ref{hdot}) that $dE/dt \propto F^2 \gamma(E) $ 
and all curves in Fig. \ref{fig2}c would overlap.  Unlike results for 
nonintegrable insulators shown in Fig. \ref{fig2}d, for integrable case the
slope of $E(t)$  increases faster than $F^2$. Although the accessible
values of $L$ do not allow for a detailed size scaling,  Fig. \ref{fig2}c  strongly suggests that the nonlinearity is not a finite--size effect.

Fig. \ref{fig2_extra} shows our central result:  $F$ dependence of the ratio $I/F$  
for $\beta_{eff} \ll 1$. In the case of nonintegrable 
insulators, $I/F$ approaches  LR $\sigma(0)$ when $F \rightarrow 0$ (dotted horizontal line), while 
weak but systematic decreasing of this ratio occurs for larger $F$. 
This is consistent with a nonmonotonic $I$--$F$ characteristic which
has been demonstrated for various tight--binding models \cite{prl2}.
%The dc current is maximal for large but finite field  and  (algebraically) decays with $F$ when $F \gg 1$. 
However, in integrable insulator the ratio $I/F$ depends on $F$ 
in a qualitatively different way and LR regime is not visible 
down to the lowest accessible fields.

\begin{figure}
\includegraphics[width=0.48\textwidth]{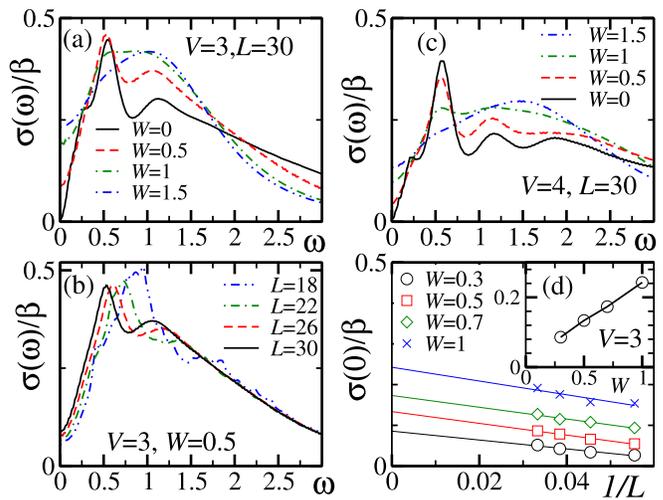}
\caption{(Color online) Equilibrium LR: $\sigma(\omega)/\beta$ in the limiting case
$\beta \rightarrow 0$ for various $W$ (panels a and c) 
and various system sizes $L$ (panel b). 
Main panel in d) shows $\sigma(0)/\beta$ vs. $1/L$ (points) while inset shows 
$W$--dependence of $\sigma(0)/\beta$ for $V=3$ extrapolated to $L \to \infty$ .}
\label{fig3}
\end{figure}

Since the difference due to integrability is most
pronounced for low electric fields, the underlying mechanism should be visible
also in the LR. Fig. \ref{fig3} shows equilibrium optical conductivity obtained within
the microcanonical Lanczos method introduced in Ref. \cite{mclm,u4}.
As already noticed for the integrable insulator ($V>2$) \cite{u4} $\sigma(\omega>0)$ 
is dominated by a finite--size dependent low frequency peak at $\omega_0 \sim 16 /L$, 
its position only weakly varying with $V$.
 But the existence and the value of $\sigma(0)$
(at any $L$) critically depends on $W$  (see Figs. \ref{fig3}a and \ref{fig3}c).
 Since $W \ne 0$ (as well as $F>0$) breaks integrability one can expect 
finite $\sigma(0)$ in the thermodynamic limit. However the $L \to \infty$ scaling
 remains delicate
%%%In particular,  $\sigma(\omega)$  has a few characteristic features, which when
%% scaled to $L \rightarrow \infty$ may give different predictions for $\sigma(0)$
as shown in Fig. \ref{fig3}b for $W=0.5$. For $\omega>\omega_{FS}\simeq 1.5$ 
all the data form a single $L$-independent curve  which could suggest with an
oversimplified extrapolation to  $L \rightarrow \infty$ that $\sigma(0)/\beta > 0.5$. 
 However, the high--$T$ sum rule  (at half--filling),
$
\int_0^{\infty} d{\rm \omega} \sigma(\omega) = \frac{\pi}{4} \beta,   
$
which is moreover $V,W$ independent, 
requires $L$--independent $\int_0^{\omega_{FS}} d{\rm \omega} \sigma(\omega)$, 
hence prevents an extrapolation to e.g. $\sigma(0)/\beta > 0.4$.
A direct scaling of
$\sigma(0)/\beta$ suggests $\sigma(0)/\beta \simeq 0.15$ for $W=0.5$ 
as shown in Fig.  \ref{fig3}d.
Analogous scaling for other $W > 0$  leads to values of $\sigma(0)$ 
decreasing and  (linearly) vanishing with $W$ (see inset in Fig. \ref{fig3}d).
In any case, the resulting $\sigma(0)$ 
is much smaller than straightforward
scaling of anomalous fluctuations for $W=0$ which would  lead, e.g. for
$V=3$, to $\sigma(0) > 0.3$ \cite{u4}. Clearly, the surprising result 
$\sigma(0)_{(W \to 0)} \to 0$ is consistent with the field dependence $I/F \to 0$ for $F \to 0$
% and explains the nonlinear $I$--$F$ characteristic 
shown in  Fig. \ref{fig2_extra}.
In like manner as  $W$,   finite $F$ as well  breaks integrability,
leading effectively to an enhancement of the response function. Consequently $I$ increases with $F$ faster than linearly.

\ \\
\textit{Discussion.} ---While our numerical results were obtained only for very high $T \gg 0$ it
is plausible that their implications  might be extended to all $T>0$, especially since 
at low temperatures strongly nonlinear $I$--$F$ characteristic is well 
established \cite{landau}.
The notion of 'ideal insulator' means vanishing of the LR
$\sigma(0)$ at arbitrary temperature  that implies at the same time 
a nonlinear   dc response  of integrable insulators 
down to the lowest electric fields. However, for
realistic (open) systems it means that the LR regime is determined 
by the mechanisms which break integrability (see discussion in Ref. \cite{afleck}), hence this regime may terminate at 
low fields. 
Here, we present two independent results which support the 
hypothesis of 'ideal insulator': nonlinear real--time current for finite driving
and LR $\sigma(\omega)$ obtained for small but finite integrability--breaking parameter $W$.
The Anderson insulator (noninteracting case when all states are localized)
has been the only well known example of an 'ideal insulator'.   
An emerging question concerns the differences between the present study
and those which suggest $\sigma(0)>0$ \cite{znid,stein}. 
A possible explanation is that  in our setup  driving
doesn't destroy the homogeneous distribution  of charge carriers, 
what may happen, e.g., in an open system driven by  a difference in chemical potentials. 
We have also studied an isolated system exactly at half--filling (microcanonical ensemble). 
Consequently, the Mott insulating phase is not destroyed (even locally) by a departure 
from  half--filling and integrability is not broken by  external leads.

Our results confirm that  the Mott insulators, in particular the
integrable ones, behave under driving fields qualitatively different  
from usual band insulators. While in the latter case thermally (as well as
photo etc.) induced carriers, i.e. electrons and holes,
are quite independent but oppositely charged quasipartilces,
in Mott insulators the corresponding doublons and 
holons are highly correlated, and moreover cannot pass each other in the
integrable 1D system (evident in the limit $V>>2$). Hence no
dc current can be induced at weak $F$ at any $T$.
On the one hand, anomalous low-frequency peak at  $\omega \propto 1/L$ clearly requires
caution in the interpretation.  On the other hand, the concept of the mean free path 
$\lambda$ would be ill defined if the $1/L$ dependence in dc response 
persists also for $L \gg \lambda$.

%It is  surprising that our results for integrable insulator are similar 
%to the breakdown of the Mott insulator \cite{landau}  
%in that  departure from linear $I-F$ dependence is visible for small rather than large driving.
%Certainly we investigate energies far above the Mott gap 
%so the underlying mechanisms are very  different. However, 
%if the present nonlinearity persists at lower temperatures  
%one may expect an interesting interplay between both  mechanisms. 

\acknowledgements

This work has been supported by the Program P1-0044 of the Slovenian Research Agency (ARRS) and
RTN-LOTHERM project. M.M. acknowledges support from the N N202052940 project of MNiSW.


\begin{thebibliography}{99}
\bibitem{kollar}
M. Kollar, F. A. Wolf, and M. Eckstein, arXiv:1102.2117;
P. Barmettler {\it et al}, New J. Phys. {\bf 12}, 055017 (2010).
\bibitem{rigol} 
M. Rigol {\it et al.}, Phys. Rev. Lett. {\bf 98}, 050405 (2007);
A.C. Cassidy, C.W. Clark, and M. Rigol, Phys. Rev. Lett. {\bf 106}, 140405 (2011).
\bibitem{u2} X. Zotos  and P. Prelov\v{s}ek,  Phys. Rev. B {\bf 53}, 983 (1996).
\bibitem{prl1} M. Mierzejewski and P. Prelov{\v{s}}ek,  Phys. Rev. Lett. {\bf 105}, 186405 (2010).
\bibitem{zotos} X. Zotos, Phys. Rev. Lett. {\bf 82}, 1764 (1999);
for an overview see
 F. Heidrich--Meisner, A. Honecker, and W. Brenig,  Eur. Phys. J. Special Topics  {\bf 151}, 135 (2007).
\bibitem{afleck} J. Sirker, R. G. Pereira, and I. Affleck,  Phys. Rev. Lett. {\bf 103}, 216602 (2009);  Phys. Rev. B {\bf 83}, 
035115 (2011).
\bibitem{u4} P. Prelov{\v{s}}ek {\it et al},  Phys. Rev. B {\bf 70}, 205129 (2004).
\bibitem{znid} T. Prosen and M. \v Znidari\v c, J. Stat. Mech.: Theory Exp. (2009), P02035;
M. \v Znidari\v c,  Phys. Rev. Lett. {\bf 106}, 220601 (2011).
\bibitem{stein} R. Steinigeweg and J. Gemmer, Phys. Rev. B {\bf 80}, 184402 (2009);
R. Steinigeweg, and R. Schnalle,   Phys. Rev. E {\bf 82}, 040103(R) (2010). 
\bibitem{sah} C. Buragohain and S. Sachdev,  Phys. Rev. B {\bf 59}, 9285 (1999).   
\bibitem{mclm} M. W. Long {\it et al}, Phys. Rev. B {\bf 68}, 235106 (2003).
\bibitem{lantime} T. J. Park and J. C. Light, J. Chem. Phys. {\bf 85}, 5870 (1986).
\bibitem{rigolnature}  M. Rigol, V. Dunjko, and M. Olshanii, Nature {\bf 452}, 854 (2008).
\bibitem{prl2} M. Mierzejewski {\it et al}, Phys. Rev. Lett. {\bf 106}, 196401 (2011); 
L. Vidmar {\it et al}, Phys. Rev. B {\bf 83}, 134301 (2011);
C. Aron, G. Kotliar, and C. Weber, arXiv: 1105.5387. 
\bibitem{landau} 
T. Oka and H. Aoki, Phys. Rev. Lett. {\bf 95}, 137601 (2005);
A. Takahashi, H. Itoh, and M. Aihara, Phys. Rev. B {\bf 77}, 205105 (2008);
N. Sugimoto, {\it et al}, Phys. Rev. B 78, 155104 (2008);
M. Eckstein, T. Oka, and P. Werner, Phys. Rev. Lett. {\bf 105}, 146404 (2010).
\end{thebibliography}
\end{document}